\documentclass[twocolumn,pre,floatfix]{revtex4}
\usepackage{graphicx}

\newcommand{\eqref}[1]{(\ref{#1})}
\newcommand{\mycaps}[1]{\makebox[0pt][s]{#1\hfill}}
\newcommand{\myvspace}{}
\newcommand{\naive}{na\"\i ve}

\newcommand{\adhoc}{\emph{ad hoc}}
\newcommand{\etal}{\emph{et al}}
\newcommand{\DebHuck}{Debye-H\"uckel}

\newcommand{\McMay}{McMillan-Mayer}
\newcommand{\nm}{\mathrm{nm}}
\newcommand{\angstrom}{\hbox{\AA}}
\newcommand{\molar}{{\mathrm{M}}}
\newcommand{\millimolar}{{\mathrm{mM}}}

\newcommand{\kB}{k} 
\newcommand{\kT}{\kB T}

\newcommand{\F}{^{\mathrm{(F)}}}
\newcommand{\X}{^{\mathrm{(X)}}}
\newcommand{\btwo}{B_2}
\newcommand{\btwohs}{B_2^{\mathrm{HS}}}
\newcommand{\hs}{_{\mathrm{HS}}}
\newcommand{\salt}{c_s}
\newcommand{\reservoir}{^{\mathrm{(R)}}}
\newcommand{\saltr}{\salt\reservoir}
\newcommand{\protein}{c_p}
\newcommand{\vp}{V_p}
\newcommand{\Q}{Q} 
\newcommand{\base}{^{(0)}}
\newcommand{\NA}{N_{\mathrm{A}}}
\newcommand{\pH}{p{\mathrm{H}}}
\newcommand{\rc}{r_c}
\newcommand{\thc}{\theta_c}
\newcommand{\nstick}{n_s}
\newcommand{\beps}{\beta\epsilon}
\newcommand{\phimin}{\phi_{\mathrm{min}}}
\newcommand{\dphi}{\delta\phi}
\newcommand{\crit}{^{\mathrm{(C)}}}
\newcommand{\pot}{\varphi}
\newcommand{\opot}{{\overline\pot}}
\newcommand{\dopot}{\Delta\pot}
\newcommand{\const}{C}
\newcommand{\phis}{\phi_s}
\newcommand{\phisF}{\phi_{s,\mathrm{F}}}
\newcommand{\phisX}{\phi_{s,\mathrm{X}}}
\newcommand{\phiF}{\phi_{\mathrm{F}}}
\newcommand{\phiX}{\phi_{\mathrm{X}}}
\newcommand{\phisr}{\phis\reservoir}
\newcommand{\mus}{\beta\mu_s}

\begin{document}

\title{Simple models for charge and salt effects in protein
crystallisation}

\author{P. B. Warren}
\affiliation{Unilever Research and Development Port Sunlight,
Bebington, Wirral, CH63 3JW, United Kingdom}

\date{June 13, 2002 --- Final version, accepted J. Phys. Cond. Mat.}

\begin{abstract}
A simple extension of existing models for protein crystallisation is
described, in which salt ions and charge neutrality are explicitly
incorporated.  This provides a straightforward explanation for the
shape of protein crystallisation boundaries, the associated scaling
properties seen for lysozyme, and can also explain much of the
salt dependence of the second virial coefficient.  The analysis has
wider implications for the use of pair potentials to understand
protein crystallisation.
\end{abstract}


\maketitle

\section{Introduction}
Protein crystallisation is of great practical importance, both as a
purification method and since high quality crystals are needed for
X-ray diffraction work \cite{CrystBook,Piazza-rev}. Under typical
conditions where crystals are obtained, protein-protein interactions
appear to be characterised by hard core repulsions with short-ranged
attractions \cite{Poon,Lek}. For instance, Rosenbaum \etal\ \cite{RZZ}
successfully collapsed the crystallisation boundaries for a number of
proteins onto the adhesive hard sphere (AHS) model\ \cite{MG,TB,HF} by
matching the second virial coefficient $\btwo$.  Whilst the AHS model
is frequently used to understand protein crystallisation, it is also
possible that crystallisation is an energetic ordering transition
driven by highly directional interactions.  A model of hard spheres
with sticky patches was introduced by Sear to capture this important
possibility \cite{Sear}.

In this paper I examine the properties of a simple model for protein
crystallisation which takes into account the effects of charge in a
very elementary way.  Such a model suggests that some generic aspects
of protein crystallisation can be explained as a straightforward
consequence of charge neutrality, at least for lysozyme which readily
undergoes crystallisation.  These various aspects are: (i) the overall
shape of the crystallisation boundary, (ii) a scaling collapse of the
crystallisation boundary noticed by Poon \etal\ \cite{PEBSS} when
plotted as a function of $\salt/\Q^2$, where $\salt$ is the NaCl
concentration and $\Q$ the protein charge; (iii) a similar scaling of
the lysozyme solubility data in Guo \etal\ \cite{GKMACW}; and (iv) a
similar scaling for $\btwo$ noticed by Egelhaaf and Poon \cite{EP}.

The appearance of $\salt/\Q^2$ in all these is very reminicent of the
Donnan membrane equilibrium \cite{Donnan}, where one finds a
contribution to $\btwo$ equal to $\Q^2/4\salt$ provided one is in the
high-salt limit $\salt\gg \Q\protein$ where $\protein$ is the protein
concentration.  The effect is well known in polyelectrolyte theory
\cite{Don1,Don2}.  This result is usually derived by supposing ideal
solution behaviour and imposing electrical neutrality on the two sides
of a semi-permeable membrane.  In the present case, I argue there is
an analagous Donnan equilibrium between protein crystals and a dilute
protein solution, with the crystal-solution interface playing the role
of the membrane. The appearance of $\salt/\Q^2$ as a scaling variable
reflects the fact that the solution (not necessarily the crystal) is
in the high-salt limit, as will be explained below.  Donnan's theory
appears to have little to do with the more conventional \McMay\
approach \cite{MM,Belloni} in which effective potentials between
proteins are invoked to explain crystallisation and the effects of
added salt, often in the context of DLVO potentials \cite{PEBSS} where
\DebHuck\ theory is used to account for the charge interactions.
However, a careful study of the relationship between the two
approaches made by Hill in the 1950s indicates both approaches should
converge on the same results in the high salt limit \cite{Hillnote}.

To apply these ideas to the protein crystallisation problem, one
constructs a free energy which enforces the charge neutrality
constraint.  Small ions can be treated as ideal solution species, but
one must at least incorporate non-ideality of the protein solution, to
allow it to form the ordered phase whch represents the protein
crystals.  The first thing to do is to set up this general theory.

\section{General theory}\label{sec-gen}
I suppose the proteins under consideration have a molecular volume
$\vp$, for example lysozyme is a charged globular protein of
approximate size $\vp = (\pi/6) \times 4.5 \times 3.0 \times
3.0\,\nm^3 = 21.2\,\nm^3$ \cite{PEBSS,BTS}, or a molar volume
$\NA\vp=12.8\,\molar^{-1}$ where $\NA$ is Avogadro's number.  If the
protein concentration is $\protein$, the effective volume fraction is
$\phi=\vp\protein$.  In the absence of charge effects, a baseline
model which incorporates a freezing transition to correspond to
protein crystallisation will be fully specified by a dimensionless
free energy density $f\base=\vp\beta F\base/V$, where $F\base$ is the
free energy of the protein solution or crystal, $V$ is the system
volume, and $\beta=1/\kT$.  There may be different branches of
$f\base$ to correspond to the fluid and ordered (crystal) phases.
Below I shall consider two possibilities for this baseline model:
Sear's model of hard spheres with sticky patches, and the AHS model.
The full free energy is obtained from the baseline model by adding in
contributions from the coions and counterions (which I suppose to be
univalent) and imposing charge neutrality.  If the added salt
concentration is $\salt$, there will be coions at a concentration
$\salt$ and counterions at a concentration $\Q\protein+\salt$.
Denoting the dimensionless free energy density for the full model by
$f$, I write
\begin{equation}
\begin{array}{r}
f(\phi,\phis)=f\base(\phi)+\phis[\log\phis-1]\hspace{6em}\\[4pt]
{}+(\Q\phi+\phis)[\log(\Q\phi+\phis)-1]
\end{array}
\end{equation}
where the last two terms are the ideal mixing terms from the coions
and counterions respectively, written using $\phis\equiv\vp\salt$ for
notational simplicity.  Unimportant constants and terms linear in
$\phi$ or $\phis$ have been dropped.  Non-ideality of the small ions
is neglected in the present treatment, although this is certainly a
refinement which should be considered for more quantitative work.

This free energy is a function of two density variables: $\phi$
and $\phis$.  A phase equilibrium such as between protein crystals and
protein solution corresponds to equality of osmotic pressure and
chemical potentials for \emph{both} components.  As a consequence
there will in general be \emph{different} values of $\phis$ in
coexisting phases, a salt repartitioning effect which is not usually
taken into account.  To solve for phase equilibria, it is useful to
transform $f$ into a semi-grand potential $h(\phi,\phisr)$ where
$\phisr\equiv\vp\saltr$ is a dimensionless salt \emph{reservoir}
concentration \cite{PBW1}.  To make this transformation note that the
salt chemical potential is
\begin{equation}
\mus=\frac{\partial f}{\partial\phis}=\log\phis+\log(\Q\phi+\phis).
\end{equation}
Now $\phis\to\phisr$ in the limit $\phi\to0$, therefore $\mus =
2\log\phisr$.  This gives
\begin{equation}
(\phisr)^2=\phis(\Q\phi+\phis)\label{doneqeq}
\end{equation}
which is the usual Donnan equilibrium result that the product of the
coion and counterion concentrations takes a constant value in all
phases including the reservoir.  Solving this gives
\begin{equation}
\phis=[(\Q^2\phi^2+4(\phisr)^2)^{1/2}-\Q\phi]/2.\label{phisreq}
\end{equation}
The semi-grand potential $h=f-\mus\phis$ and the first two
derivatives with respect to $\phi$ at constant $\phisr$
are, after a few lines of calculus,
\begin{eqnarray}
&&h=f\base+\Q\phi\log(\Q\phi+\phis)-(\Q\phi+2\phis),\label{heq}\\
&&\partial h/\partial\phi=\partial f\base\!/\partial\phi
+\Q\log(\Q\phi+\phis),\label{dheq}\\
&&\partial^2\!h/\partial\phi^2=\partial^2\!f\base\!/\partial\phi^2
+\Q^2/(\Q\phi+2\phis).\label{d2heq}
\end{eqnarray}
The advantage of this transformation is that $h$ is effectively a
one-component free energy and can be treated accordingly.  To use
these results, one should remember to substitute for $\phis$ from
Eq.~\eqref{phisreq}.  For example, the osmotic pressure follows from
$\vp\beta\Pi=\phi(\partial h/\partial\phi)-h$, ie
\begin{equation}
\Pi=\Pi\base+\kT(\Q\protein+2\salt).\label{osmeq}
\end{equation}
This shows explicitly the small ions behaving ideally, at a total
concentration $\Q\protein+2\salt$.

What is obvious from these results is that there is a cross-over in
behaviour at $\Q\protein\sim\salt$ or salt concentrations of the order
$\Q\phi/\vp$.  The \emph{difference} in protein volume fraction
between the solution and crystal is often $\Delta\phi\sim0.5$.
Putting $\Q\sim10$, and $\vp\sim10\,\molar^{-1}$, this corresponds to
a cross-over salt concentration $\sim0.5\,\molar$.  If the salt
concentration is much less than this, there will be a large osmotic
pressure difference between the crystal and the solution due to the
counterions, having the effect of narrowing the coexistence gap.  This
is the basic reason why the coexistence boundary occurs at salt
concentrations of this order of magnitude in this model.  It is a
much larger cross-over salt concentration than intuition might have
suggested based on experiences for colloidal systems (eg $\Q\sim10^3$,
$\vp\sim10^{6}\,\nm^3$ gives $\salt\sim10^{-3}\,\molar$), but then
globular proteins are much smaller than colloids.

On the other hand, typical protein solutions on the crystallisation
boundary have $\phi\sim0.05$, corresponding to a cross-over salt
concentration $\sim50\,\millimolar$.  For salt concentrations much
larger than this, the effects of salt and charge are subsumed into a
scaling variable $\phisr/\Q^2$.  Since $\phisr$ is more or less the
salt concentration in the solution if the protein concentration is
small, this is a natural explanation for the scaling properties described
in the introduction.  In the next sections, I place these arguments on
a firm footing, starting with discussions of the second virial
coefficient and the high-salt scaling of solubility, since these are
not dependent on any particular baseline model.

\begin{figure}
\begin{center}
\mycaps{(a)}\includegraphics{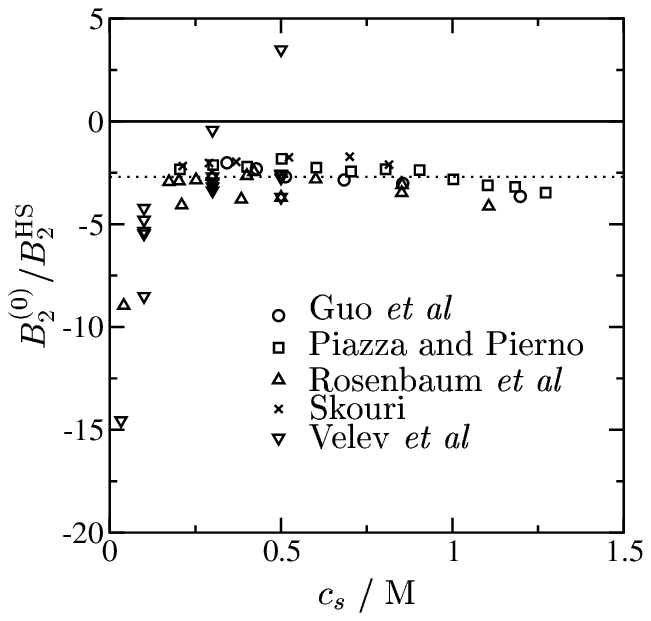}
\myvspace
\mycaps{(b)}\includegraphics{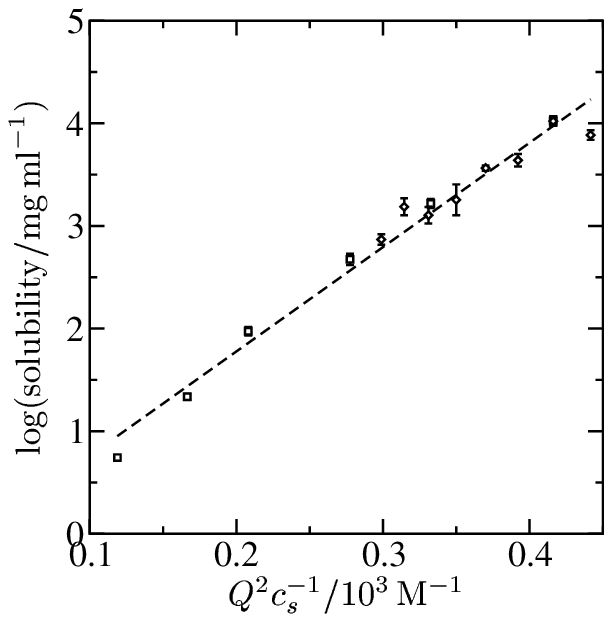}
\end{center}
\caption[?]{(a) $\btwo\base/\btwohs$ for lysozyme, where
$\btwo\base=\btwo-\Q^2/4\salt$, as a function of salt concentration
$\salt$ \cite{EP}.  The dotted line is the average,
Eq.~\eqref{b2baseq}, for all data with $\salt>0.25\,\molar$.  (b)
Solubility data for lysozyme from Guo \etal\ \cite{GKMACW}.  The
dashed line is the least-squares fit to the data,
Eq.~\eqref{bestfiteq} in the text.\label{fig-b2}}
\end{figure}

\section{Second virial coefficient}\label{sec-b2}
Any theory for the free energy of a protein solution contains a
prediction for the second virial coefficient.  For the present theory,
$\btwo$ can be obtained from the osmotic pressure result
Eq.~\eqref{osmeq} (recalling that Eq.~\eqref{phisreq} should be used
for $\phis$), or directly by inspection from Eq.~\eqref{d2heq}.
Either way, as advertised in the introduction, $\btwo = \btwo\base +
\Q^2/4\salt$ where $\btwo\base$ is the contribution from the baseline
model.  One can prove the same result holds in light scattering
determination since charge density fluctuations are suppressed in the
long wavelength limit.  We can hope that the baseline model is
insensitive to the net charge and salt concentration, so that
$\btwo\base$ is independent of $\Q$ and $\salt$, but is there any
evidence for this?  For lysozyme, Egelhaaf and Poon have collected
experimental data on $\btwo$ from the available literature
\cite{EP}.  This data is used to construct Fig.~\ref{fig-b2}(a)
which shows $\btwo - \Q^2/4\salt$ normalised by the value expected if
lysozyme proteins behaved as hard spheres, namely $\btwohs =
4\vp\approx85\,\nm^3$ (this is close to $\btwohs=82.3\,\nm^3$ used in
Ref. \cite{RKRZ}) Inspecting Fig.~\ref{fig-b2}(a), it appears there is
indeed a data collapse to an approximate plateau for $\salt \agt
0.25\,\molar$, although a downwards trend of 40\% or so can be
detected at high salt and there may also be a problem if the protein
charge is too small (the `rogue' point with $\btwo\base>0$ is for
$\Q=3.6$).  There is considerable variation in the data, even for
measurements at supposedly identical state points, which reflects the
experimental difficulties in obtaining $\btwo$.  Averaging over all
results in the plateau region in Fig.~\ref{fig-b2}(a) arrives at
\begin{equation}
\btwo\base\approx(-2.7\pm0.2)\times\btwohs.\label{b2baseq}
\end{equation}
This result can be used as a constraint for any particular baseline
model for lysozyme.  Furthermore, the approximate constancy of
$\btwo\base$ as $\Q$ and $\salt$ are varied suggests that it is at
least reasonable to make the baseline model independent of $\Q$ and
$\salt$, as is assumed in the remainder of this paper.

\section{High salt scaling behaviour}
In the high salt limit, a simple result can be obtained for the
crystallisation boundary (equivalently the protein solubility) by
assuming that the baseline model is independent of salt and charge,
the density in the crystal remains constant, and the protein
concentration in the solution phase becomes very small so that the
solution can be treated as ideal.  These features are certainly seen
for the two specific baseline models discussed below.  The
simplifications amount to treating $\partial f\base\!/\partial\phi$ in
Eq.~\eqref{dheq} as a constant independent of $\Q$ and $\salt$ in the
protein crystal, and taking the dilute limit $\partial
f\base\!/\partial\phi\to\log\phiF$ in the solution where $\phiF$ is
the protein concentration in the solution phase.  Equating protein
chemical potentials in the two phases results in
\begin{equation}
\begin{array}{r}
\log\phiF+\Q\log(\Q\phiF+\phisF)\hspace{6em}\\[4pt]
{}=\mathrm{constant}+\Q\log(\Q\phiX+\phisX)
\end{array}
\end{equation}
where $\phisF$ and $\phisX$ are the salt concentrations in the
solution and crystal phases respectively, and $\phiX$ is the protein
volume fraction in the crystal.  The terms in this can be expanded
assuming that $\phiF\ll1$, $\phisF\approx\phisr$, $\phisX\gg\Q\phiX$,
and making use of Eq.~\eqref{phisreq} to get
\begin{equation}
\log\phiF\approx\mathrm{constant}+\frac{\Q^2\phiX}{2\phisF}.\label{hisalteq}
\end{equation}
The prediction is that the logarithm of the protein solubility should
be proportional to the square of the protein charge, and inversely
proportional to the salt concentration in the solution.
Fig.~\ref{fig-b2}(b) shows that such a law is indeed satisfied for the
solubility data for lysozyme from Guo \etal\ \cite{GKMACW}.  The best
fit line in Fig.~\ref{fig-b2}(b) is
\begin{equation}
\begin{array}{r}
\displaystyle
\log\Bigl(\frac{\mathrm{solubility}}{\mathrm{mg}\,\mathrm{ml}^{-1}}\Bigr)
=(-0.25\pm0.15)\hspace{6em}\\[-4pt]
\displaystyle
{}+(0.010\pm0.001)\times
\frac{\Q^2}{\salt/\molar}.
\end{array}
\label{bestfiteq}
\end{equation}
The predicted slope from Eq.~\eqref{hisalteq} is $\phiX/2\vp$.  If we
take $\phiX \approx 0.5$ as a reasonable estimate of the protein
volume fraction in the crystal, then $\phiX/2\vp \approx 0.02\,\molar$
which around twice the measured slope.  However there are a
number of omitted effects in the present theory which could account
for the discrepancy.  Nevertheless the observation that the logarithm
of the protein solubility is inversely proportional to the inverse
salt concentration is apparently quite
common \cite{fn1}.  Also note that solubility
$\sim e^{\Q^2}$ indicates a very strong dependence on the protein
charge.  This is a natural explanation for the dramatic decrease in
solubility around the isoelectric point described in
Ref.~\cite{CrystBook}.

\section{Two specific baseline models}
For lower salt concentrations, the protein solubility is not small and
the above treatment needs refinement.  I now consider the minimal
extension to the basic model that takes into account protein
non-ideality by specifying a particular baseline model.  In this way,
the advantage of an analytic transformation to the semi-grand
potential $h$ in Eqs.~\eqref{heq}--\eqref{d2heq} is retained.  The
baseline model should encompass both the fluid and crystal phases, and
have a second virial coefficient consistent with the analysis above.
One such suitable model has been devised by Sear.

\subsection{Sear's model}
Sear's model comprises hard spheres (HS) with `sticky patches'
\cite{Sear}.  The HS diameter is $\sigma$, chosen such that
$\pi\sigma^3/6=\vp$ (ie $\sigma\approx 3.4\,\nm$ for lysozyme).  There
are $\nstick$ sticky patches per sphere which associate in pairs and
are characterised by a range $\rc>\sigma$, an angular width $\thc$,
and a depth $\epsilon$.  The free energy of the fluid (F) phase (I
reproduce only the essential details of the model here) is
\begin{equation}
f\base_{\mathrm{F}}(\phi)=f\hs(\phi)+\phi\nstick[\log p+(1-p)/2].
\end{equation}
In this $f\hs$ is the HS fluid free energy and $p$ is the proportion of
non-bonded sites, solving $(1-p)/p^2=k\phi g\hs$ where
\begin{equation}
k=6(\rc/\sigma-1)(1-\cos\thc)^2e^{\beps}
\end{equation}
is a dimensionless bond association constant.  The association
equilibrium includes an enhancement factor, $g\hs$, for the HS pair
correlation function at contact.  The second virial coefficient in
this model is given by
\begin{equation}
\btwo\base=\btwohs-(\nstick/2)\,k\vp.\label{pahsb2eq}
\end{equation}
It is apparent that the fluid phase properties are completely
determined by $k$ and $\nstick$.  The model predicts fluid-fluid phase
separation for sufficiently large values of $k$, and Table
\ref{tab-crit} gives the critical point for $\nstick=4$--8.  Note that
the second virial coefficient at the critical point in this model
provides an marked counterexample to the observation of
Vliegenthart and Lekkerkerker that $\btwo\crit / \btwohs \approx -1.5$
for a wide variety of other models \cite{VL}.  Whether fluid-fluid
phase separation is metastable in this model depends on the actual
values of $\rc$, $\thc$ and $\beps$, and examples of both are given by
Sear.

\begin{table}
\caption[?]{Fluid-fluid critical points for Sear's model showing
critical parameter values $\phi\crit$, $k\crit$ and $\btwo\crit$ for
several values of $\nstick$.\label{tab-crit}}
\begin{ruledtabular}
\begin{tabular}{cccc}
$\nstick$ & $\phi\crit$ & $k\crit$ & $\btwo\crit/\btwohs$ \\
\colrule
4 & 0.09 & 16.8  & $-$7.38 \\
5 & 0.12 &  7.02 & $-$3.39 \\
6 & 0.15 &  3.90 & $-$1.93 \\
8 & 0.21 &  1.77 & $-$0.77 \\
\end{tabular}
\end{ruledtabular}
\end{table}

\begin{figure}
\begin{center}
\mycaps{(a)}\includegraphics{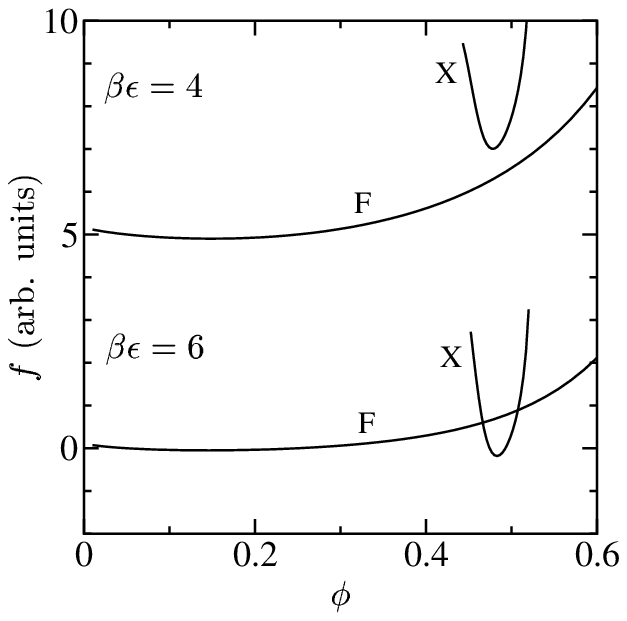}
\myvspace
\mycaps{(b)}\includegraphics{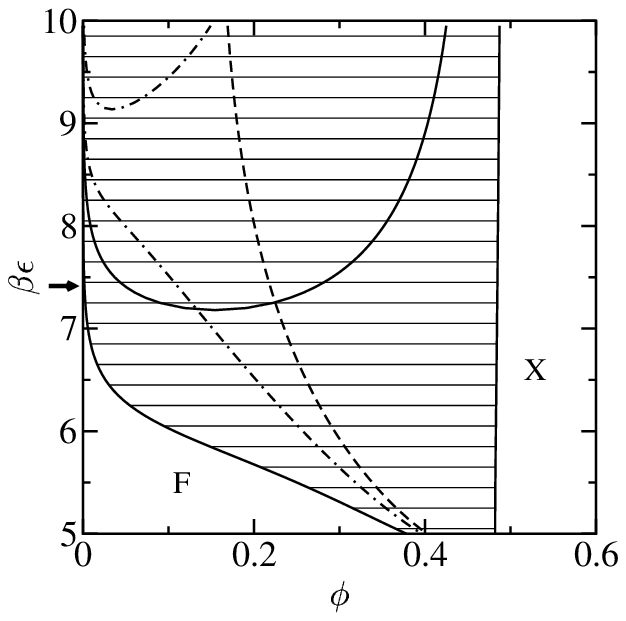}
\end{center}
\caption[?]{(a) Free energies for baseline model illustrating
metastability of crystal branch (X) with respect to fluid branch (F)
for $\beps\alt4.8$. (b) Phase behaviour in $(\phi,\beps)$ plane for
full model: solid lines are for the baseline model ($\Q=\saltr=0$),
dashed line is for $\Q=10$ and $\saltr=0$, and chained lines are for
$\Q=10$ and $\saltr = 0.2\,\molar$.  Other parameters are $\nstick=6$,
$\rc=1.05\sigma$, $\thc=0.45$.\label{fig-frengy}}
\end{figure}

Sear provides a cell model for the protein chemical potential in an
ordered cubic phase, arguing that osmotic pressure is not important.
My approach here is slightly different.  I use the same cell model to
specify the free energy and take into account the osmotic pressure.
The results are not essentially different from Sear's results,
although the analysis does point up an important property of Sear's
model (Fig.~\ref{fig-frengy}(a) below).  Following Sear, the free energy
per protein in the crystal (X) is
\begin{equation}
\begin{array}{r}
\displaystyle
\frac{f\base_{\mathrm{X}}(\phi)}{\phi}=\const
-3\log\Bigl(\frac{a}{\sigma}-1\Bigr)\hspace{6em}\\
\displaystyle
{}-\log\Bigl(\frac{\thc^3}{\pi^2}\Bigr)
-\frac{\nstick}{2}\,
\beps\,w\Bigl[\frac{(\phi-\phimin)}{\dphi}\Bigr]
\end{array}
\label{pahsfseq}
\end{equation}
where $a/\sigma=(6\phi/\pi)^{-1/3}$ is the unit cell size relative to
the HS diameter.  The constant in this is $C=\log(\vp/\sigma^3)$,
provided that $f\hs\sim\phi(\log\phi-1)$ in the fluid phase as
$\phi\to0$.  This free energy is appropriate for $\phi\agt\phimin$
where $\phimin=\pi(\sigma/\rc)^3/6$ is the volume fraction around
which the bonds in the crystal become dissociated.  As $\phi$
decreases past $\phimin$ the last term in Eq.~\eqref{pahsfseq}
vanishes rapidly.  To implement this I have introduced an \adhoc\
cut-off function $w[(\phi-\phimin)/\dphi]$ in the last term, where
$\dphi$ sets the rate at which the cut-off operates.  For the present
calculations I take $w[x]=1/(1+e^{-x})$ and $\dphi=0.01$.  This
cut-off function represents the way in which the short-range
attraction potential falls off with distance in the model.  The actual
details may shift the phase boundaries but are not important for the
broad picture.

Although several values of the parameters in the model were examined,
I only report in detail here on calculations for $\nstick=6$,
$\rc=1.05\sigma$ and $\thc=0.45$ (so the range of the attraction is
$\rc-\sigma\approx 2\,\angstrom$ and the angular width about
$26^\circ$).  These values were chosen to give quite good agreement
with the crystallisation boundaries for lysozyme in the present model.
Interestingly, in a separate analysis Curtis \etal\ \cite{CBP} also
conclude $\nstick=6$--8 is appropriate, and Oki \etal\ \cite{OMKC}
identify three `macrobonds' between lysozyme molecules from
crystallographic data, again corresponding to $\nstick=6$ contacts per
protein.

Fig.~\ref{fig-frengy}(a) shows the fluid and crystal free energies for
$\nstick=6$, $\rc=1.05\sigma$ and $\thc=0.45$, for two values of $\beps$.
It is clear that there is a certain minimum value $\beps\approx4.8$
below which the crystal is \emph{metastable} with respect the
fluid. This is an important contrast with the AHS model (next subsection)
in which the ordered phase is present at $\beps\to0$.  Crystallisation
in Sear's model is essentially an \emph{energetic} transition to an
ordered phase dominated by the directional interactions (compare
`energetic fluid' concept introduced by Louis \cite{Louis}) and not a
continuation of the \emph{entropic} HS freezing transition.  The
position of the sticky patches controls the crystal structure, and as
suggested by Sear, this may explain the relative ease or difficulty of
crystallising various proteins.

Fig.~\ref{fig-frengy}(b) shows the phase behaviour for the model for
the chosen parameter set.  The solid lines in Fig.~\ref{fig-frengy}(b)
are for the baseline model and include representative tie-lines.  As
$\beps$ increases, the fluid-crystal phase transition widens (the
re-entrant fluid phase expected at larger $\phi $ is not shown).
There is a metastable fluid-fluid phase separation for
$\beps\agt7.18$.

The effect of charges and added salt is obtained by inserting the
baseline model into the general formalism in section~\ref{sec-gen}.
The dashed line in Fig.~\ref{fig-frengy}(b) shows the fluid-crystal
phase boundary at $\Q=10$ in the absence of salt.  The transition has
been markedly narrowed and the fluid-fluid transition moves to such a
high value of $\beps$ that it is no longer on the diagram.  Repeating
the calculation for $\saltr=0.2\,\molar$ obtains the chained lines in
Fig.~\ref{fig-frengy}(b).  The fluid-cystal phase boundary is
intermediate between the zero salt limit and the baseline model, and
the metastable fluid-fluid transition has reappeared in the diagram
for $\beps\agt9.14$.  Finally as $\saltr\to\infty$, the phase
boundaries all move back to coincide with the baseline model.  Thus
the effect of charge in the model is to strongly suppress existing
phase transitions in the absence of added salt.  Adding salt weakens
and eventually destroys the effect.  The salt concentration required
to do this is $\salt\sim\Q\protein$, as discussed already in
section~\ref{sec-gen}.

\begin{figure}
\begin{center}
\mycaps{(a)}\includegraphics{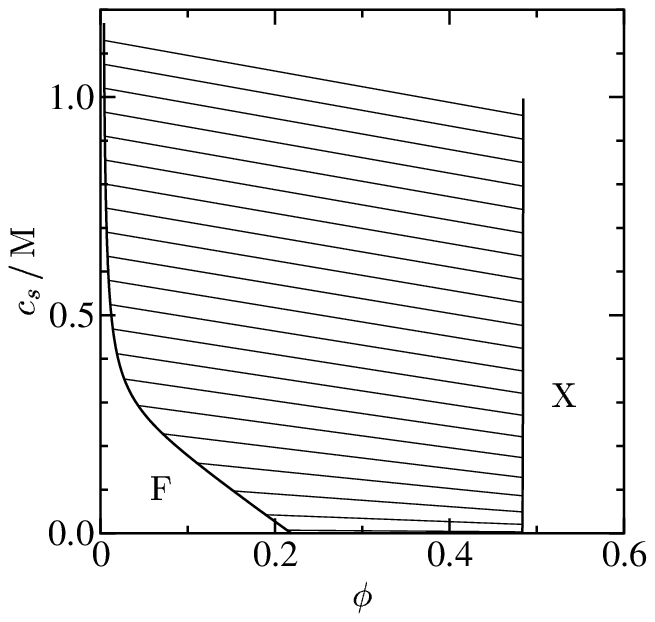}
\myvspace
\mycaps{(b)}\includegraphics{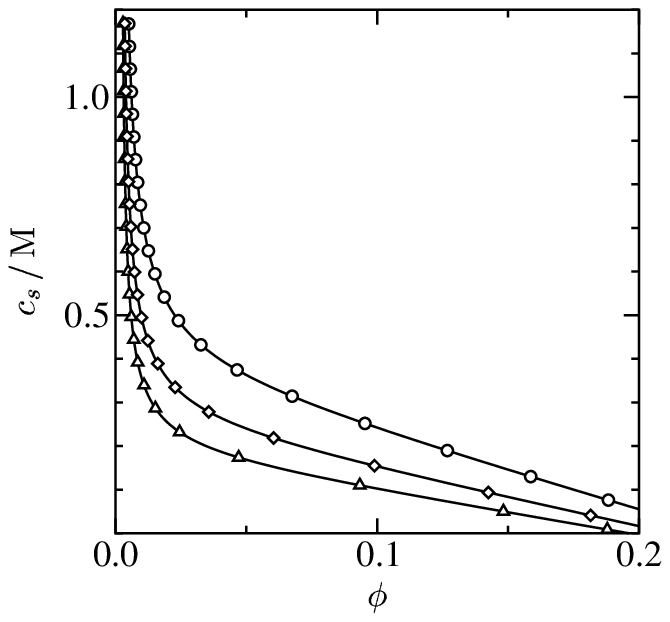}
\end{center}
\caption[?]{(a) Phase behaviour in $(\phi,\salt)$ plane for Sear's
model for $\Q=10$.  (b) Fluid-crystal binodals for $\Q=8.0$
(triangles), 9.4 (diamonds) and 11.4 (circles). Other parameters are
$\nstick=6$, $\rc=1.05\sigma$, $\thc=0.45$,
$\beps=7.4$.\label{fig-salt}}
\end{figure}

I now set $\beps=7.4$, marked by an arrow in Fig.~\ref{fig-frengy}(b),
so that $\btwo\base/\btwohs=-2.7$ reproduces the value calculated in
section~\ref{sec-b2} above \cite{fn2}.  
Fig.~\ref{fig-salt}(a) shows a typical phase
diagram for the full model at this value of $\beps$, as a function of
added salt.  Salt reservoir concentrations have been converted back to
actual salt concentrations.  The sloping tie-lines indicate a salt
repartitioning effect in which the coion concentration is enhanced in
the more dilute phases.  In this simple model I have not taken into
account excluded volume in the dense phases, so in reality the salt
repartitioning would be more marked (see below for a further
discussion on this).  In this representation we can clearly see how
adding salt broadens the fluid-crystal coexistence.  Metastable
fluid-fluid phase separation does not appear until salt concentrations
$\agt2.3\,\molar$ for this value of $\beps$.  Note that the shape of
the fluid-crystal binodal (which I interpret to be the same thing as
the crystallisation boundary) is in agreement with common experience.
This is a natural but non-trivial consequence of charge neutrality in
the present model.

\begin{figure}
\begin{center}
\mycaps{(a)}\includegraphics{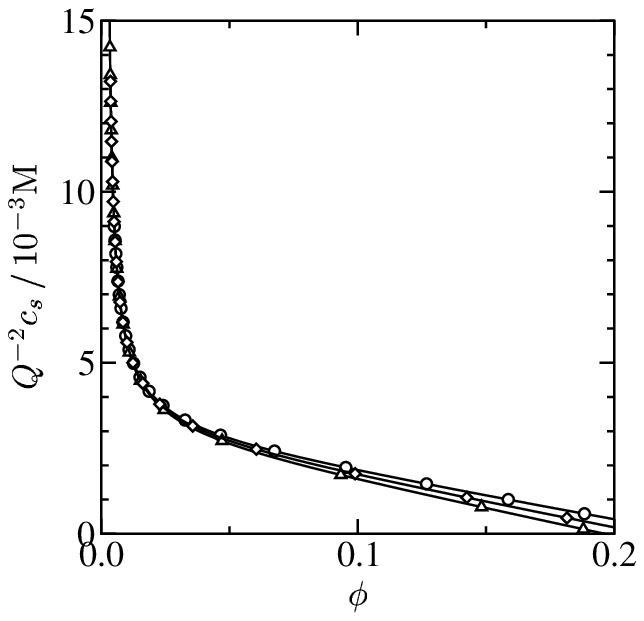}
\myvspace
\mycaps{(b)}\includegraphics{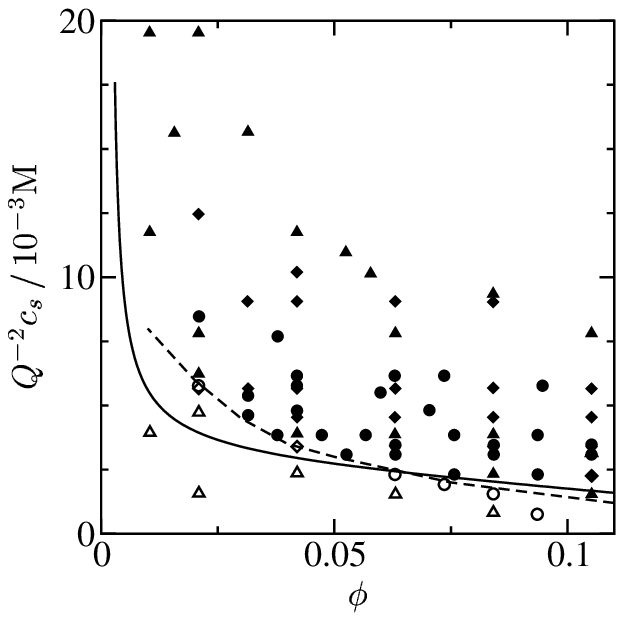}
\end{center}
\caption[?]{ (a) Data of Fig.~\ref{fig-salt}(b) replotted in
$(\phi,\Q^{-2}\salt)$ plane.  (b) Experimental data of Poon \etal\
\cite{PEBSS}, for $\Q=8.0$ (triangles), 9.4 (diamonds) and 11.4
(circles).  Filled (open) symbols indicate the occurence
(non-occurence) of crystals. The dashed line is the approximate
experimental crystallisation boundary.  The solid line as the mean
crystalisation boundary from (a).\label{fig-expt}}
\end{figure}

This calculation is now repeated for $\Q=8.0$, 9.4 and 11.4, which are
the three values of lysozyme charge examined experimentally by Poon
\etal.  Fig.~\ref{fig-salt}(b) shows the crystallisation boundary as a
function of added salt, for the three values of $\Q$.  Clearly, the
higher the charge, the more the phase transition is suppressed.
Finally, these same phase boundaries are replotted in
Fig.~\ref{fig-expt}(a) as a function of the scaling variable
$\salt/\Q^2$.  In this representation, the curves all collapse to lie
on approximately the same quasi-universal crystallisation boundary.
The scaling collapse is robust: if the calculations are repeated for
different parameter values, the quasi-universal crystallisation curve
moves up or down but a similar scaling collapse is always obtained.

The mean crystallisation boundary from Fig.~\ref{fig-expt}(a) is shown
in Fig.~\ref{fig-expt}(b), which now includes the experimental data
from Poon \etal.  There is reasonable agreement on the location of the
quasi-universal crystallisation boundary, although it appears there is
always some discrepancy between the shape of the theoretical and
experimental boundaries at low $\phi$ which recalls the discrepancy in
the slope of the high-salt law in Fig.~\ref{fig-b2}(b).  It is worth
re-emphasising that the present model is constrained to have the
correct $\btwo$, at least for $\salt\agt0.25\,\molar$.  The absence of
a metastable fluid-fluid phase separation agrees with Poon \etal\ who
observe that no such phase transition occurs at the temperature of the
experiments ($22.5\,{}^\circ\mathrm{C}$).  However, a temperature
decrease of just a few percent in the model, so that $\beps=7.7$ for
example, is sufficient to bring fluid-fluid phase separation to
accessible salt concentrations in the range $0.5$--$1.0\,\molar$.
This is in accordance with the observations of Muschol and Rosenberger
\cite{MR} (but additional caution is required since it is unlikely the
only effect of temperature is through the value of $\beps$).  In the
scaling representation of Fig.~\ref{fig-expt}(a) such metastable
fluid-fluid binodals also collapse to a quasi-universal curve---this
is a prediction of the theory that would be interesting to test
experimentally.

The scaling collapse in Fig.~\ref{fig-expt} occurs because the effects
of charge and salt in fluid free energy are effectively combined into a
single scaling variable $\salt/\Q^2$, at salt concentrations in the
vicinity of the crystallisation boundary.  To show this is not just a
feature of Sear's model, I now turn briefly to the more commonly
studied AHS model.

\subsection{Adhesive hard sphere model}
The adhesive hard sphere (AHS) model is also a suitable baseline model
for the general theory in section \ref{sec-gen}.  In this model, hard
spheres interact with a short range isotropic attractive potential.
As noted by Rosenbaum \etal\ \cite{RZZ}, the phase behaviour is
largely insensitive to the details of the potential provided the
second virial coefficient is used as the effective temperature axis.
My approach to the baseline model here is closely based on that of
Noro \etal\ \cite{NKF} who investigated the effects of long range
forces on the AHS model.

For the fluid phase, I use Barboy's treatment of Baxter's analytic
theory \cite{Barboy}.  In Baxter's theory \cite{Baxter}, the
attractive potential is characterised by a `stickiness' parameter
$\tau$ which is related to the second virial coefficient
by
\begin{equation}
{\btwo\base}/{\btwohs}=1-({4\tau})^{-1}.\label{baxteq}
\end{equation}

The crystal phase is expected to be FCC since this has the greatest
density of intersphere contacts.  As the stickiness is switched off
($\tau\to\infty$ or $\beps\to0$) the fluid-crystal phase transition
goes over into the usual HS freezing transition, which is a contrast
to Sear's model.  The AHS ordered phase has always proved rather more
difficult to treat analytically than the fluid phase, and approaches
have ranged through density functional calculations by Marr and Gast
\cite{MG}, and Tejero and Baus \cite{TB}, to detailed simulation
studies \cite{HF}.  Here I use the cell model of Daanoun \etal\
\cite{DTB} for the crystal phase free energy, assuming the attractive
part of the AHS potential is $-\epsilon(r/\sigma)^{-n}$ with $n\gg1$.
The cell model free energy is almost identical to that already used in
Eq.~\eqref{pahsfseq} with $\nstick=12$ for the FCC structure.  The
difference is that there is no restriction on orientation so the term
$-\log(\thc^3/\pi^2)$ is absent, and the role played by the \adhoc\
cut-off function is taken over by the functional form of the actual
short range potential, ie $w[x]\to(a/\sigma)^{-n}$.  Like the choices
for the cut-off function in the previous section, the actual value of
$n$ may shift the phase boundaries but does not change the broad
picture.  For the purposes of the present calculation I set $n=30$.

\begin{figure}
\begin{center}
\mycaps{(a)}\includegraphics{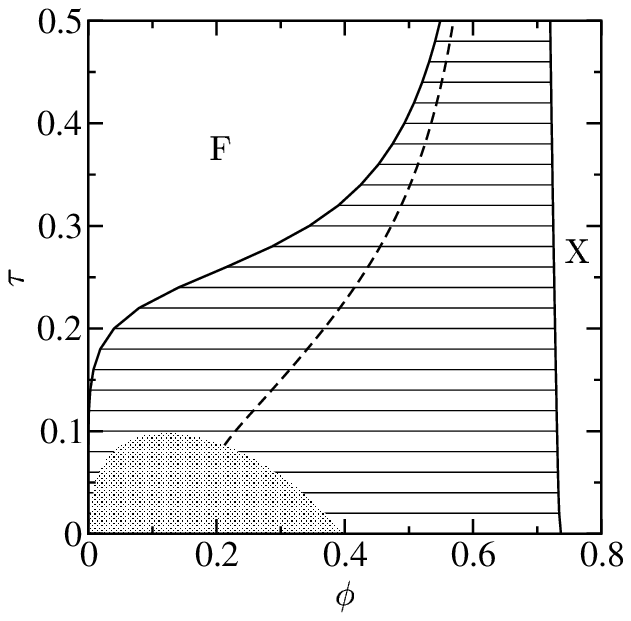}
\myvspace
\mycaps{(b)}\includegraphics{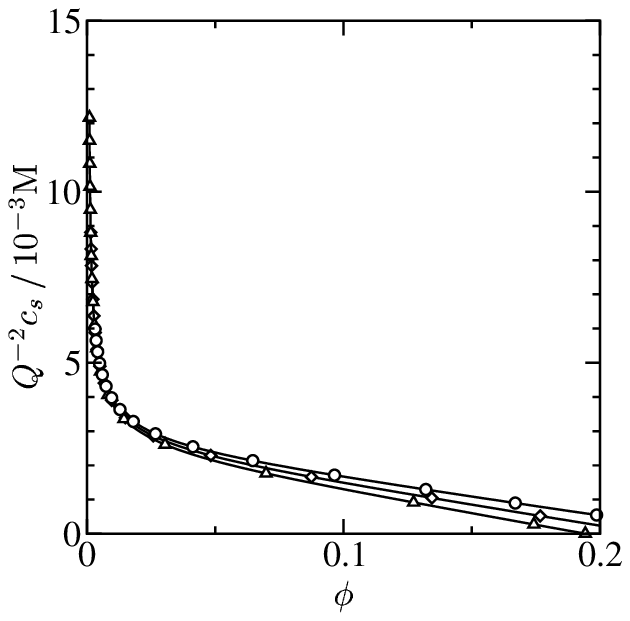}
\end{center}
\caption[?]{(a) Phase behaviour in $(\phi,\tau)$ plane for the
adhesive hard sphere model with $n=30$: solid lines are for baseline
model with $\Q=\saltr=0$, dashed line is for $\Q=10$ and $\saltr=0$.
The shaded region is the `danger zone' Eq.~\eqref{dangereq}. (b)
Fluid-crystal binodals in $(\phi,\Q^{-2}\salt)$ plane for $\Q=8.0$
(triangles), 9.4 (diamonds) and 11.4 (circles); other parameters
are $n=30$, $\tau=0.1$.\label{fig-ahs}}
\end{figure}

A link between the cell model for the crystal phase and the
Baxter-Barboy treatment of the fluid phase can be made by matching
$\btwo$ of Eq.~\eqref{baxteq} with the exact expression
\begin{equation}
{\btwo\base}/{\btwohs}=1+
3\int_1^\infty\! dx\,x^2\,[1-\exp(-\beps x^{-n})].
\end{equation}
Fig.~\ref{fig-ahs}(a) shows the phase behaviour for the model.  The
solid line in this figure is the fluid-crystal phase boundary for the
baseline model, and the dashed line is the corresponding boundary for
$\Q=10$ with no salt.  Similar to Fig.~\ref{fig-frengy}(b), the effect
of charge is to narrow the fluid-crystal transition.  As salt is added
(not shown here), the boundary moves back towards that of the baseline
model.

There is a complication that arises for Baxter's solution for the AHS
fluid phase free energy.  Baxter's theory involves the solution of a
quadratic equation.  If
\begin{equation}
\tau<\frac{(12\phi+6\phi^2)^{1/2}-6\phi}{6(1-\phi)}\label{dangereq}
\end{equation}
then this quadratic equation has complex roots and Baxter's theory
becomes inadmissable.  Normally this `danger zone', shown shaded in
Fig.~\ref{fig-ahs}(a), is happily hidden within the fluid-crystal two
phase region.  In the charged version though, it emerges into the
single phase fluid region.  For this reason one cannot choose $\tau <
(2-\surd2)/6 \approx 0.098$, the maximum of Eq.~\eqref{dangereq}, and
unfortunately this excludes $\tau = 0.067$ which would match
$\btwo\base/\btwohs = -2.7$ from section \ref{sec-b2}.

To demonstrate the scaling collapse for this model, I therefore choose
$\tau=0.1$ (equivalent to $\btwo\base/\btwohs=-1.5$ \cite{VL}.)  I
repeat the calculations of the previous section to obtain the
crystallisation boundaries (fluid-crystal binodal) as a function of
added salt, for $\Q=8.0$, 9.4 and 11.4 as used previously.
Fig.~\ref{fig-ahs}(b) shows these boundaries plotted using the scaling
variable $\salt/\Q^2$.  Again the curves collapse to a quasi-universal
crystallisation boundary, similar to those in Fig.~\ref{fig-expt}(a).
This result confirms that the scaling collapse is a common feature for
two different baseline models for the phase behaviour.

\section{Salt repartitioning and interface structure}\label{sec-don}
These ideas have some additional interesting physical implications for
the equilibrium between crystals and solution which are explored in
this section.  The first implication concerns salt repartitioning.  In
the simple theory of section \ref{sec-gen}, the product of coion and
counterion concentrations is a constant as in Eq.~\eqref{doneqeq}.
Each phase has to be electrically neutral, so that a phase enriched in
protein is also enriched in counterions, and consequently
\emph{depleted} in coions (Donnan common ion effect).  This explains
the slope of the tie-lines in Fig.~\ref{fig-salt}(a) for example.  We can
ask: is there any independent evidence for this phenomenon?  Let me
preface the answer to this by some cautionary remarks.  Deviations
from the simple Donnan equilibrium result may arise from three
sources.  Firstly excluded volume effects mean one should really
consider the small ion concentrations in the available free volume.
This will be an important consideration in the crystal phase where the
volume fraction occupied by the protein is $\sim50$\%.  Secondly,
deviations from ideal solution behaviour can be expected at high salt
concentrations---such deviations are usually absorbed into `activity
coefficients'.  Thirdly there may be significant specific ion effects,
such as seen in the Hoffmeister series.

\begin{table}
\caption[?]{Lysozyme salt repartitioning data from Palmer \etal\
\cite{PBG}.  Solution is $0.855\,\molar$ NaCl / $0.2\,\molar$ NaAc
buffer ($pH=4.5$).  Ion concentrations in the third column are
calculated by multiplying those in the second column by $1/(1-\phi)$,
where $\phi=0.64$ is the crystal volume fraction.  The charge
unaccounted for in the crystal, assuming $\Q=11.4$, is $0.05\,\molar$
or $\alt3\%$ of the total charge.\label{tab-salt}}
\begin{ruledtabular}
\begin{tabular}{rrrr}
conc / M               & solution & crystal & crystal free volume \\
\colrule
Lysozyme$^{\Q+}$       &          & 0.056   &      \\
Ac$^-$                 &   0.2    &         &      \\
Na$^+$                 &   1.055  & 0.19    & 0.53 \\
Cl$^-$                 &   0.855  & 0.78    & 2.2  \\
\colrule
Na$^+{}\times{}$Cl$^-$ &   0.9    &         & 1.2  \\
\end{tabular}
\end{ruledtabular}
\end{table}

Repartitioning of salt was studied in detail by Vekilov \etal\
\cite{VMTSR}.  They found non-uniform salt concentrations inside
protein crystals, but consider this is likely to be an effect of
impurities or growth kinetics.  The present theory only addresses the
equilibrium repartitioning of salt in perfect lysozyme crystals
though.  Fortunately, some relevant equilibrium results for salt
repartitioning can be found in recent experiments by Morozova \etal\
\cite{MKLEBSM} on cross-linked lysozyme crystals in contact with salt
solutions.  For example, Table \ref{tab-salt} shows concentrations of
Na$^+$ and Cl$^-$ in solution and in the crystal from the early work
of Palmer \etal\ \cite{PBG}, cited by Morozova \etal.
There is marked repartitioning, for example the coion concentration in
the crystal free volume is about half that in the external solution.
The product of the two ion concentrations is approximately constant
though (final row in Table~\ref{tab-salt}), provided the protein
excluded volume is taken into account.  Detailed calculations by
Morozova \etal\ for their more recent data also take into account
activity coefficients and give excellent agreement between theory and
experimental results: they conclude ``\dots for small ions capable of
penetrating into the crystal channels [the] electrostatic (Donnan)
potential controls the equilibrium internal concentration of ions in
just the same way as in polyelectrolyte gels'' \cite{MKLEBSM}.  By way
of contrast, the same work also suggests Br$^-$ has a significant
specific interaction with lysozyme.

The second implication concerns the electrical structure of the
interface between crystal and solution phases.  In a Donnan membrane
equilibrium \cite{Donnan,Don1,Don2}, an electrostatic potential
difference, the \emph{Donnan potential}, develops between the two
compartments on either side of the membrane.  In the present theory
for protein crystallisation, it is still true that a difference in the
mean electrostatic potential develops between the crystals and the
solution phase.  Whether this should also be called a Donnan
potential, a \emph{Galvani potential}, or perhaps something else, can
be debated \cite{Fisher,Sparney}.

At any rate, the potential difference is readily calculated given the
salt repartitioning.  If $c_+$ and $c_-$ are the coion and counterion
concentrations respectively, then $c_\pm = \saltr\,\exp[\mp\beta
e\opot]$, where $\opot$ is the mean electrostatic potential in the
phase of interest measured relative to the salt reservoir (this result
can also be used to recover Eq.~\eqref{doneqeq}).  The difference in
mean potential between fluid (F) and crystal (X) phases,
$\dopot=\opot\X-\opot\F$, is $\dopot=(\beta
e)^{-1}\log(\salt\F/\salt\X)$.  For example, for the salt
concentrations in Table~\ref{tab-salt}, $\salt\F=1.055\,\molar$,
$\salt\X=0.53\,\molar$ (in the free volume), and therefore
$\dopot\approx17\,\mathrm{mV}$.  The potential difference arises
because an electrical double layer is formed at the crystal-solution
interface and is intimately connected with salt repartitioning.  The
details of this will be discussed in a forthcoming paper \cite{SW}.
Another consequence of the Donnan potential is that the $\pH$ in the
crystal will be $0.434\dopot$ higher than in the solution (where
$0.434=\log_{10}e$). If the solution $\pH$ is below the isoelectric
point, as is often the case for lysozyme, one might expect this to lower the
charge per protein in the crystal although it is unlikely that the
charge can be determined with any accuracy from such a \naive\ calculation.

Pair-potential theories miss both the effects of salt repartitioning
and the potential difference between the two phases, which are
essentially many-body phenomena \cite{Langmuir,PBW2}).  Whilst this is
not a problem at high salt where Hill's mapping between the \McMay\ /
\DebHuck\ pair potential approach and Donnan's method goes through
\cite{Hillnote}, the effects can be significant at lower salt
concentrations.  The importance of the effects can be gauged by
comparing the magnitude of $\dopot$ to the thermal energy $\kT$.  If
$\dopot\agt\kT/e$ then there may be significant errors introduced by
using a single effective pair potential (such as a DLVO potential) for
both the crystal and the solution phases.  On the other hand, if
$\dopot\ll\kT/e$, the use of pair potentials cannot be criticised on
these grounds. It can easily be shown that $\dopot\agt\kT/e$
corresponds to the by now familiar $\salt\alt\Q\protein$, in other
words salt concentrations smaller than the protein charge
density in the crystal, which is typically $0.5\,\molar$.

\section{Conclusions}
There are several conclusions from the present work.  The first
concerns the lysozyme / NaCl system.  The quasi-universal
crystallisation boundary observed by Poon \etal\ \cite{PEBSS} can be
fit reasonable well by the present version of Sear's model with
$\sim6$ sticky contacts per molecule, similar to the conclusions of
other workers \cite{CBP,OMKC}.  The scaling collapse for $\btwo$
noticed by Egelhaaf and Poon \cite{EP} is here attributed to a
contribution $\Q^2/4\salt$ added to a `bare' $\btwo\base$ which is
approximately \emph{independent} of salt concentration above about
$0.25\,\molar$.  In this system therefore, NaCl appears to be acting
as an \emph{indifferent electrolyte} in the sense that it does not
seem to exhibit specific ion effects.  This appears to be confirmed by
the experiments of Morozova \etal\ \cite{MKLEBSM} discussed in the
preceeding section.  Other salts, for example NaBr, may not behave the
same way of course.

The other conclusions are general ones. Firstly, a simple extension of
existing models to incorporate salt ions and charge neutrality
provides a straightforward explanation for the shape of protein
crystallisation boundaries and the associated scaling properties seen
for lysozyme.  Even if the present theory proves inadequate to
describe the experiments in quantitative detail, this is surely a
robust observation.  Secondly, there are a large number of effects
which have been omitted from the present theory, but which could be
incorporated with some additional numerical effort, such as excluded
volume effects, non-ideality of the salt ions, and other electrostatic
correlation efects.  Finally, the twin phenomena of salt
repartitioning and the concomitant appearance of a significant Donnan
potential difference discussed in the previous section are essentially
many-body effects, and are not captured in simple pair-potential
theories.


{\hfill oOo\hfill}

For discussions and correspondence, I thank S. U. Egelhaaf,
M. E. Fisher, H. N. W. Lekkerkerker, M. G. Noro, W. C. K. Poon and
R. P. Sear.  Finally, it is a great pleasure to dedicate this paper to
Peter Pusey on the occasion of his 60th birthday.


\begin{thebibliography}{10}
\bibitem{fn1}
Ref.~\cite{CrystBook} suggests alternatively that the logarithm of the
protein solubility should be proportional to ionic strength, but this
is a poorer fit to the solubility data of Guo \etal\ \cite{GKMACW} than
Eq.~\eqref{bestfiteq}.
\bibitem{fn2}
Oki \etal\ \cite{OMKC} suggest `macrobond' energies of the order $50\,
\mathrm{kcal}\, \mathrm{mol}^{-1} \approx 80\,\kT\gg\beps$ but this
estimate is for the bare macrobond energy and does not include
re-solvation of the protein surface.
\end{thebibliography}
\end{document}